\documentclass[
reprint,           
superscriptaddress,
amsmath,           
amssymb,           
prd,               
notitlepage,       
longbibliography,  
floatfix,          
]{revtex4-1}

\usepackage{tensor}     
\usepackage{graphicx}   
\usepackage[
colorlinks=true,        
citecolor=blue,         
linkcolor=blue,         
urlcolor=blue           
]{hyperref}             
\usepackage{bm}         
\usepackage{xcolor}     

\newcommand{\newc}{\newcommand*} 
\newc{\figurewidth}{3.2in}
\newc{\xbar}{\bar{x}}
\newc{\rhoeq}{\rho_{\rm{eq}}}
\newc{\zeq}{z_{\rm{eq}}}
\newc{\la}{\lambda}
\newc{\tla}{\tilde{\la}}
\newc{\dt}{\delta}
\newc{\Dt}{\Delta}
\newc{\vj}{\vec{j}}
\newc{\vl}{\vec{l}}
\newc{\hx}{\hat{x}}
\newc{\hy}{\hat{y}}
\newc{\bj}{\bm{j}}
\newc{\mJ}{\mathcal{J}}
\newc{\mP}{\mathcal{P}}
\newc{\ga}{\gamma}
\newc{\Msun}{M_\odot}
\newc{\app}{\approx}
\newc{\av}[1]{\langle #1 \rangle}
\newc{\eq}[1]{Eq.~\eqref{#1}}
\newc{\al}{\alpha}
\newc{\Xstar}{X_{\ast}}
\newc{\seq}{\sigma_{\rm{eq}}}
\newc{\fpbh}{f_{\rm{pbh}}}
\newc{\VT}{\langle VT \rangle}

\def\p{\partial}

\def\({\left(}
\def\){\right)}
\def\[{\left[}
\def\]{\right]}

\def\e{\begin{equation}}
\def\q{\end{equation}}
\def\m{\begin{eqnarray}}
\def\n{\end{eqnarray}}

\begin{document}

\title{Merger Rate Distribution of Primordial-Black-Hole Binaries}

\author{Zu-Cheng Chen}
\email{chenzucheng@itp.ac.cn} 
\affiliation{CAS Key Laboratory of Theoretical Physics, 
Institute of Theoretical Physics, Chinese Academy of Sciences,
Beijing 100190, China}
\affiliation{School of Physical Sciences, 
University of Chinese Academy of Sciences, 
No. 19A Yuquan Road, Beijing 100049, China}

\author{Qing-Guo Huang}
\email{huangqg@itp.ac.cn}
\affiliation{CAS Key Laboratory of Theoretical Physics, 
Institute of Theoretical Physics, Chinese Academy of Sciences,
Beijing 100190, China}
\affiliation{School of Physical Sciences, 
University of Chinese Academy of Sciences, 
No. 19A Yuquan Road, Beijing 100049, China}
\affiliation{Center for Gravitation and Cosmology, College of Physical Science and Technology, Yangzhou University, Yangzhou 225009, China}
\affiliation{Synergetic Innovation Center for Quantum Effects 
and Applications, Hunan Normal University, Changsha 410081, China}

\date{\today}
\begin{abstract}

Up to now several gravitational-wave events from the coalescences of black hole binaries have been reported by LIGO/VIRGO, and imply that black holes should have an extended mass function. We work out the merger rate distribution of primordial-black-hole binaries with a general mass function by taking into account the torques by all primordial black holes and linear density perturbations. In the future, many more coalescences of black hole binaries are expected to be detected, and the one-dimensional and two-dimensional merger rate distributions will be crucial for reconstructing the mass function of primordial black holes.

\end{abstract}

\pacs{???}

\maketitle


It is believed that primordial black holes (PBHs) could have formed in the early universe from the collapse of large density fluctuations \cite{Hawking:1971ei,Carr:1974nx,Carr:1975qj}. On the other hand, one of the challenges for the fundamental physics is the understanding of the nature of dark matter (DM). Among a large variety of models, the speculation that the DM be composed totally or partially by PBHs has attracted much attention, especially since the discovery of black hole coalescence by LIGO \cite{Abbott:2016blz}, because the coalescence of PBH binaries can be the candidates for the observed gravitational-wave events \cite{Sasaki:2016jop,Bird:2016dcv}.

Up to now, there are several gravitational-wave events from binary black hole (BBH) mergers reported by LIGO and VIRGO collaborations: GW150914 ($36_{-4}^{+5}M_\odot$, $29_{-4}^{+4}M_\odot$) \cite{Abbott:2016blz}, GW151226 ($14.2_{-3.7}^{+8.3}M_\odot$, $7.5_{-2.3}^{+2.3}M_\odot$) \cite{Abbott:2016nmj}, GW170104 ($31.2_{-6.0}^{+8.4}M_\odot$, $19.4_{-5.9}^{+5.3}M_\odot$) \cite{Abbott:2017vtc}, GW170608 ($12_{-2}^{+7}M_\odot$, $7_{-2}^{+2}M_\odot$) \cite{Abbott:2017gyy}, GW170814 ($30.5_{-3.0}^{+5.7}M_\odot$, $25.3_{-4.2}^{+2.8}M_\odot$) \cite{Abbott:2017oio}, as well as a less significant candidate LVT151012 ($23_{-6}^{+18}M_\odot$, $13_{-5}^{+4}M_\odot$) \cite{TheLIGOScientific:2016qqj,TheLIGOScientific:2016pea}. These events indicate that the black holes should have an extended mass function. In fact, the generic initial conditions of PBH formation also suggest that the PBH mass should also extend over a wide range.

In literature there are two main paths for the PBH binary formation. One is formed in the early Universe \cite{Sasaki:2016jop,Nakamura:1997sm,Ali-Haimoud:2017rtz} and another is formed in the late Universe \cite{Bird:2016dcv,Ali-Haimoud:2017rtz,Nishikawa:2017chy} respectively, and the former generically makes the dominant contribution to the PBH merger rate. However, the mass function of PBHs is usually assumed to be monochromatic in \cite{Sasaki:2016jop,Nakamura:1997sm,Ali-Haimoud:2017rtz,Bird:2016dcv,Nishikawa:2017chy}. Recently, the merger rate of PBHs with an extended mass function is investigated in \cite{Raidal:2017mfl,Kocsis:2017yty}, but only the tidal force from the PBH closest to the center of mass of the PBH binary is taken into account in \cite{Raidal:2017mfl} and a flat mass function of PBHs over a small mass range is considered in \cite{Kocsis:2017yty}. 

In this paper we consider the torques by all PBHs and linear density perturbations, and calculate the merger rate distribution for the PBH binaries with a general mass function. In the next decades, many more BBH mergers are expected to be detected and will provide much better information about the black hole mass function and may finally help us to answer what is the origin of black holes and how the black hole binaries are formed in our Unvierse.

The probability distribution function (PDF) of PBHs $P(m)$ is normalized to be 
\m
\int_0^\infty P(m)dm=1, 
\n
and the abundance of PBHs in the mass interval $(m, m+dm)$ is given by 
\m
f P(m)dm, 
\n
where 
$f$ is the total abundance of PBHs in non-relativistic matter. In this paper, for convenience, the PBH mass is in unit of $M_\odot$. The fraction of PBHs in cold DM is related to $f$ by $f_{\rm{pbh}}\equiv \Omega_{\rm{pbh}}/\Omega_{\rm{cdm}} \approx f/0.85$. 
We introduce a cross-grained discrete PDF, namely  
\m
\int P(m)dm=1\ \rightarrow \ \sum_{m_{\rm{min}}\leq m_i \leq m_{\rm{max}}} P_i \Dt \simeq 1, 
\n
where $P(m_i) \rightarrow P_i$ is the binned PDF and 
$dm_i \rightarrow \Dt$ denotes the resolution of PBH mass. Roughly speaking, $f P_i \Dt\equiv f_i \Dt$ is taken as the abundance of PBHs with mass $m_i$.
At matter-radiation equality the total energy density of matter is 
\m
\rhoeq =\Omega_m \rho_{\rm{crit}} (1+z_{\rm{eq}})^3, 
\n
and then the average distance ${\bar x}_i$ between two PBHs with mass $m_i$ is 
\m
\xbar_i=\({3\over 4\pi} {m_i\over \rhoeq f_i \Dt}\)^{1/3} 
\n
which depends on both mass $m_i$ and its abundance $f_i \Dt$. We need to stress that the number densities for PBHs with different masses can be quite different from each other, and in principle there is no well-defined number density for the over-all PBHs. 
The average distance $\langle x_{ij}\rangle$ between two neighboring PBHs with different masses  $m_i$ and $m_j$ is estimated as follows 
\m
\langle x_{ij}\rangle=  \(\xbar_i^{-3}+\xbar_j^{-3}\)^{-1/3}
    = \mu_{ij}^{1/3} \xbar_{ij}, 
\n
where 
\m
\mu_{ij}&=& {2m_im_j f_{b}\over m_{b}(f_jm_i+f_im_j)},\label{mu}\\
\xbar_{ij}^3&=&{3\over 8\pi} {m_{b}\over \rhoeq f_{b} \Dt}, 
\label{xbar}
\n
and 
\m
f_{b}&=&f_i+f_j,\\
m_{b}&=&m_i+m_j. 
\n 
The above formula are valid for $m_i\neq m_j$, and can be generalized to cover $m_i=m_j=m$ if we take $P(m_i)=P(m_j)=P(m)/2$. 
From now on, for simplicity, we omit the subscript `$_{ij}$' unless it is necessary.

In order for the formation of PBH binary, the two neighboring PBHs necessarily decouple from the background expansion and form a bound system. 
In Newtonian approximation, the equation governing the evolution of the 
proper separation $r$ of the BH binary with masses $m_i$ and $m_j$ along the axis of motion takes the form 
\m\label{eom1}
  \ddot{r} - \( \dot{H} + H^2 \) r + \frac{m_b}{r^2} \frac{r}{|r|} = 0, 
\n 
where the dot denotes the derivative with respect to the proper time. In this paper we work in the geometric units $G=c=1$. 
Defining $\chi \equiv r/x$, we re-write Eq.~\eqref{eom1} as follows 
\m 
  \chi'' + \frac{s h' + h}{s^2 h} \(s \chi' - \chi \) + \frac{1}{\la}
    \frac{1}{\(sh\)^2} \frac{1}{\chi^2} \frac{\chi}{|\chi|} = 0, 
    \label{chi}
\n 
where $x$ is the comoving separation between these two PBHs, primes denote the derivative with respect to scale factor $s$ which is normalized to be unity at equality, and $h(s)\equiv H(s)/\({8\pi\over 3}\rhoeq\)^{1/2}=\sqrt{s^{-3}+s^{-4}}$.
Here the dimensionless parameter $\la$ is 
\m 
  \la = \frac{8 \pi \rhoeq x^3}{3 m_b} = {X\over f_b\Dt},
\n 
where 
\m
  X\equiv {x^3/\xbar^3}, 
\n
and $\xbar$ is given in Eq.~\eqref{xbar}. The solution of Eq.~\eqref{chi} in \cite{Ali-Haimoud:2017rtz} implies that the decoupling before the equality if $\lambda<1$ and the semi-major axis $a$ of the formed binary is given by 
\m 
  a \approx 0.1 \la x = \frac{0.1}{f_b \Dt} \frac{x^4}{\xbar^3}
    ={0.1 \xbar \over f_b\Dt} X^{\frac{4}{3}}. 
      \label{axis}
\n 

Without the tidal force from other PBHs and density perturbations, these two PBHs will just head-on collide with each other. However, the tidal force will provide an angular momentum to prevent this system from direct coalescence. For simplicity, we introduce a dimensionless angular momentum $j$ defined by 
\m
j\equiv \ell/\sqrt{m_b a}=\sqrt{1-e^2}, 
\n
where $\ell$ is the angular momentum per unit reduced mass and 
$e\in[0,1]$ is the eccentricity. 
In order to estimate the initial orbital parameters for the PBH binary in which two PBHs can have different masses, we generalize the method in \cite{Ali-Haimoud:2017rtz}. Here we do not plan to repeat all of the calculations in \cite{Ali-Haimoud:2017rtz}, but only highlight the key results. 
The local tidal field is related to the Newtonian potential $\phi$ by $T_{ij}=-\p_i\p_j \phi$ which exerts a perturbative force per unit mass $\bm{F}=\bm{T}\cdot \bm{r}$. 
Supposing that the initial comoving separation of the binary is small relative to the the mean separation, this tidal force does not significantly affect the orbit of binary, but it produces a torque which yields 
\m
\bm{\ell}=\int dt\ \bm{r}\times [\bm{T}\cdot \bm{r}]. 
\n
Since the tidal field generated by other PBHs and density perturbation in the radiation-domination era goes like $s^{-3}$, $\bm{T}\simeq s^{-3} \bm{T}_{\rm{eq}}$ and then 
\m 
  \bj \approx x^3 \hx \times \[\frac{\bm{T}_{\rm{eq}}}{m_b} \cdot \hx \], 
\n 
where $\hx$ is the unit vector along $\bm{x}$ and $\bm{T}_{\rm{eq}}$ 
is the local tidal field at equality. 
The tidal field generated by a PBH with mass $m_l$ at a comoving separation $y\gg x$ is given by 
\m 
  T^{ij}_{eq} = m_l \frac{3 \hy^i \hy^j - \dt^{ij}}{y^3},
\n 
and then 
\m 
  \bm{j} \approx 3 \frac{m_l}{m_b} \frac{x^3}{y^3} 
    \(\hx \cdot \hy\) \(\hx \times \hy\).
\n
Similar to \cite{Ali-Haimoud:2017rtz}, following \cite{Chandrasekhar:1943ws}, the two-dimensional PDF of $j$ from torques by all other PBHs is given by 
\m 
  \frac{dP}{d^2 j} = \lim_{V \to \infty}\int \frac{d^2 k}{\(2\pi\)^2} e^{i \bm{k} \cdot \bj} 
       \prod_{l} {\cal I}_l^{N_l},
\n 
where $N_l=n_l V$ is total number of PBHs with mass $m_l$, 
\m 
  {\cal I}_l = \int_V \frac{d^3 y}{V} \exp\[-3 \frac{m_l}{m_b} i
    \frac{x^3}{y^5} y_{||}\, \bm{k} \cdot \bm{y}_{\perp}\],
\n 
and $y_{||}\equiv \bm{y}\cdot \hx$ and $\bm{y}_{\perp}\equiv \hx\times \bm{y}$. After a tedious computation, we find 
\m
  \lim_{V \to \infty} {\cal I}_l^{N_l} = e^{-{4\pi\over 3} {m_l\over m_b} n_l x^3 k}. 
\n
Since $m_l n_l=\rho_l$ is the energy density of PBHs with mass $m_l$, $\sum_l \rho_l=\rho_{\rm{pbh}}=f \rhoeq$ and then 
\m\label{prob1}
  \frac{dP}{dj} = j \int k dk J_0(kj) e^{-j_X k},
\n 
where 
\m
j_X=0.5{f\over f_b\Dt} X
\n
which encodes the torques by all other PBHs. Integrating over Eq.~\eqref{prob1} gives 
\m\label{Pj}
  \left. j \frac{dP}{dj} \right\vert_X = \mP\(j/j_X\), \quad
    \mP(\ga) = \frac{\ga^2}{\(1 + \ga^2\)^{3/2}},
    \label{pj}
\n
where $\gamma=j/j_X$. 
In addition, the variance of $\bm{j}$ due to the torques by density perturbations is 
\e
\langle j^2 \rangle^{1/2}\approx 0.5 {8\pi\over 3} {\sigma_{\rm{eq}}\rhoeq \over m_b}x^3=0.5 {\sigma_{\rm{eq}}\over f_b\Dt} X, 
\q
where $\sigma_{\rm{eq}}\equiv \langle \delta_{\rm{eq}}^2 \rangle^{1/2}$ is the variance of density perturbations of the rest of DM on scale of order ${\cal O}(10^0\sim10^3) M_\odot$ at equality. Taking into account both the torques by all of other PBHs and density perturbations, the characteristic value of $j_X$ in Eq.~\eqref{pj} reads 
\e
j_X\approx 0.5 \(f^2+\sigma_{\rm{eq}}^2\)^{1/2} {X\over f_b\Dt}. 
\label{jx}
\q

After the formation of PBH binary, the orbit of these two binary PBHs shrinks due to the gravitational waves, and the coalescence time is given by, \cite{Peters:1964zz},  
\m 
  t = \frac{3}{85} \frac{a^4}{m_i m_j m_b} j^7.
\n 
Taking into account Eq.~\eqref{axis}, the dimensionless angular momentum goes like 
\m 
  j(t; X) = \( \frac{85}{3} \frac{t m_i m_j m_b (f_b\Dt)^4}
      {\(0.1\xbar\)^4 X^{16/3}} \)^{1/7}.
\n  
Assuming that PBHs possess a random distribution, the probability distribution of the separation $x$ between two nearest PBHs with mass $m_i$ and $m_j$ and without other PBHs in the volume of ${4\pi \over 3}x^3$ becomes 
\m 
 \frac{dP}{d{\tilde X}} = e^{- {4\pi \over 3}x^3 n_T}=e^{- {\tilde X}\cdot {4\pi \over 3}\langle x_{ij} \rangle^3 n_T}, 
 \label{dpX}
\n 
where ${\tilde X}\equiv x^3/\langle x_{ij} \rangle^3=X/\mu$, $\mu$ is given in Eq.~\eqref{mu}, and $n_T\equiv f\rho_{\rm{eq}}\int_0^\infty {P(m)\over m}dm$ \cite{nt}. 
Therefore we have 
\e
  \frac{d^2 P}{d{\tilde X} dt} = \frac{1}{7 t} e^{- {\tilde X}\cdot {4\pi \over 3}\langle x_{ij} \rangle^3 n_T}\, \mP(\ga_X),
  \quad \ga_X \equiv \frac{j(t; X)}{j_X}, 
\q 
where $j_X$ is given in Eq.~\eqref{jx}. The probability distribution of the time of merger becomes 
\e 
  \frac{dP}{dt}  
      = \frac{\mu^{-1}}{7 t} \int d{X} e^{- {X\over \mu}\cdot {4\pi \over 3}\langle x_{ij} \rangle^3 n_T} \mP(\ga_X),
      \label{dP}
\q
and the comoving merger rate at time $t$ reads 
\m
R_{ij}(t)\equiv {dN_{\rm{merger}}\over dtdV}=\rho_m^0 \min\(\frac{f_i \Dt}{m_i}, \frac{f_j \Dt}{m_j}\) {dP\over dt}, 
\n
where $\rho_m^0\simeq 4\times 10^{19}\, \Msun \text{Gpc}^{-3}$ is the matter density at present. Since ${\cal P}(\gamma_X)$ has a sharp peak at 
\e
X_*(t)\approx 0.032 \({t\over t_0}\)^{3\over 37} f_b \Delta (f^2+\sigma_{\rm{eq}}^2)^{-{21\over 74}} (m_i m_j)^{3\over 37} m_b^{-{1\over 37}}, 
\q
if ${X_*\over \mu}\cdot {4\pi \over 3}\langle x_{ij} \rangle^3 n_T\ll 1$ \cite{xmu}, 
the comoving merger rate at time $t$ becomes 
\m
R_{ij}(t)= {\cal R}_{ij}(t) \Delta^2, 
\n
where  
\m\label{calR}
{\cal R}_{ij}(t)&\approx& 3.9\cdot 10^6\times  \({t\over t_0}\)^{-{34\over 37}} f^2 (f^2+\sigma_{\rm{eq}}^2)^{-{21\over 74}} \nonumber \\
&\times&  \min\(\frac{P(m_i)}{m_i}, \frac{P(m_j)}{m_j}\) \({P(m_i)\over m_i}+{P(m_j)\over m_j}\) \nonumber \\
&\times& (m_i m_j)^{{3\over 37}} (m_i+m_j)^{36\over 37},
\n
which can be interpreted as the comoving merger rate density in unit of Gpc$^{-3}$ yr$^{-1}$, and PBH masses are in unit of $M_\odot$. 
Again, we want to remind readers that $P(m_i)=P(m_j)=P(m)/2$ for $m_i=m_j=m$.  
If $P(m)/m=$ constant, $\tilde{\alpha} \equiv -(m_i+m_j)^2\p^2 \ln {\cal R}_{ij}/\p m_i \p m_j=36/37$, which is consistent with \cite{Kocsis:2017yty}. However, for a general mass function, $\tilde{\alpha}$ can be quite different from $36/37$. 

\begin{figure}[htbp!]
\centering
\includegraphics[width = 0.48\textwidth]{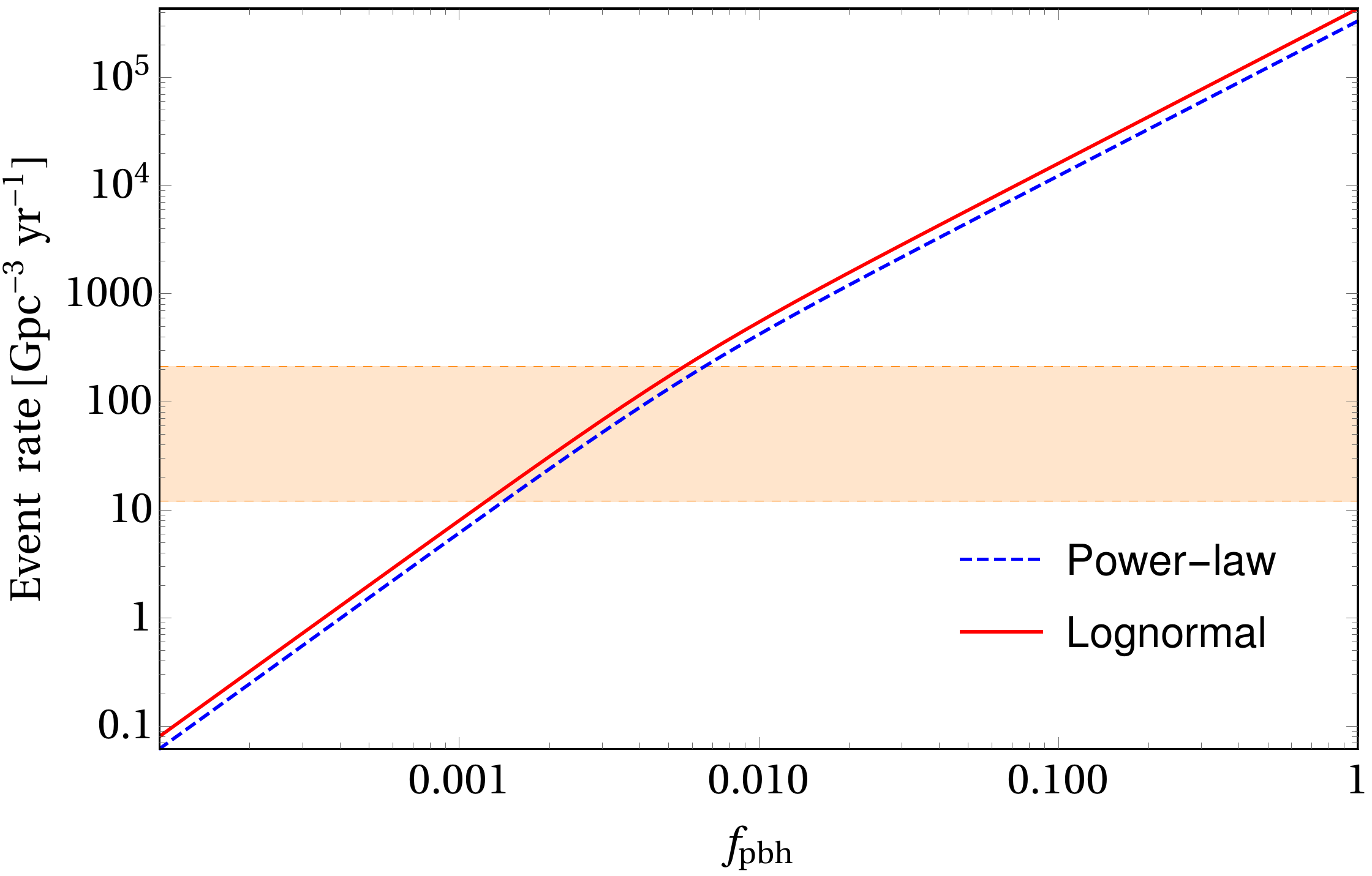}
\caption{\label{merger} 
The merger rate of PBH binaries at present with $m_1,\ m_2\geq 5M_\odot$ and $m_1+m_2\leq 100M_\odot$. 
The blue dotted and red solid lines correspond to the 
\textit{power-law} PDF ($M=5M_\odot$ and $\alpha=1.6$)
and \textit{lognormal} PDF ($m_c=15M_\odot$ and $\sigma=0.6$), respectively. 
}
\end{figure}
\begin{figure}[htbp!]
\centering
\includegraphics[width = 0.48\textwidth]{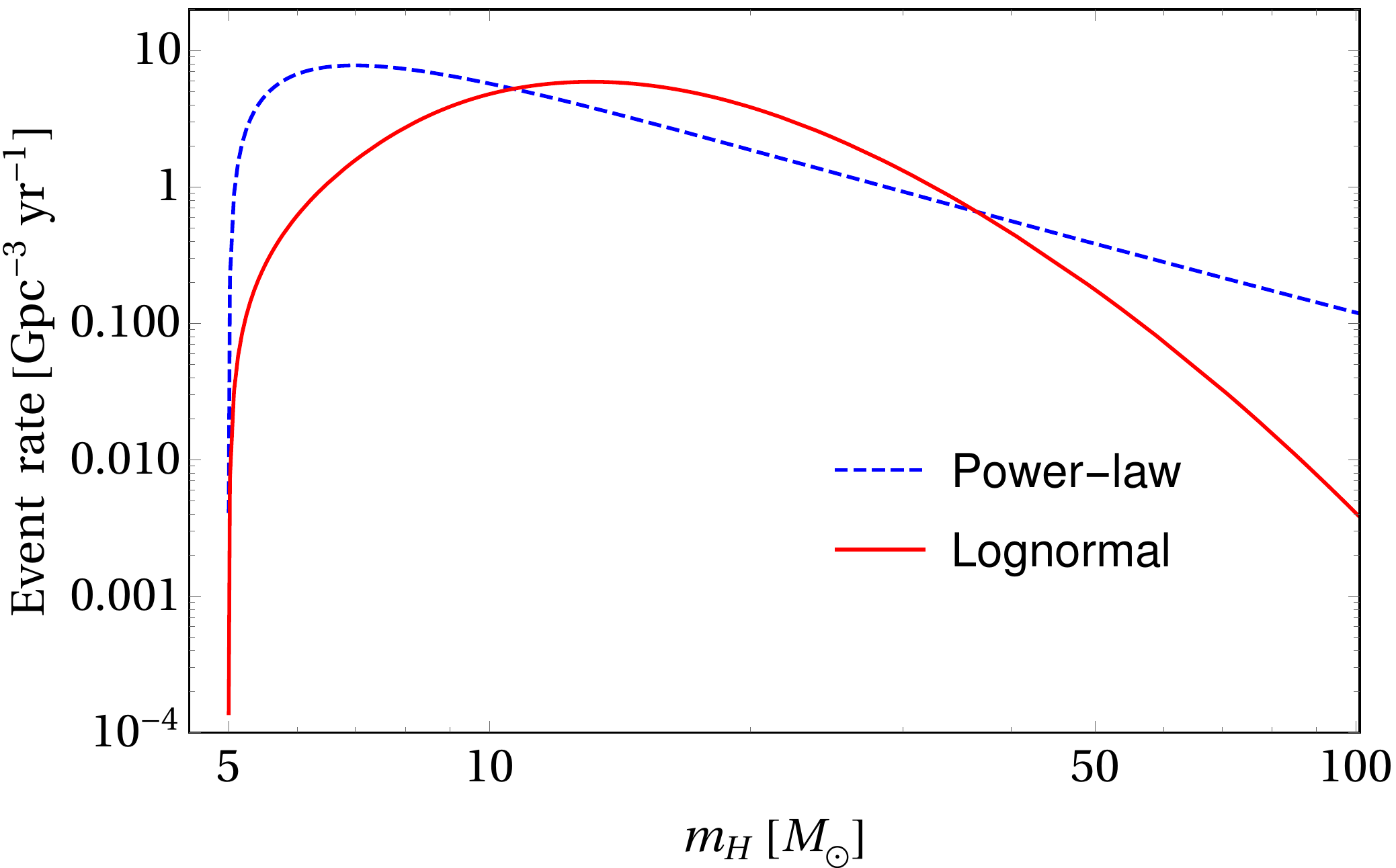}
\caption{\label{mergerheavy}
The 1D merger rate distribution, where $m_H$ is the mass of heavier BH in the binary and the mass of lighter BH is integrated over from $5M_\odot$ to $m_H$. 
The blue dotted and red solid lines correspond to the \textit{power-law} 
PDF ($M=5M_\odot$ and $\alpha=1.6$) with $\fpbh =4.3\times 10^{-3}$
and \textit{lognormal} PDF ($m_c=15M_\odot$ and $\sigma=0.6$) with $\fpbh = 3.7\times 10^{-3}$, respectively.
}
\end{figure}
\begin{figure}[htbp!]
\centering
\includegraphics[width = 0.48\textwidth]{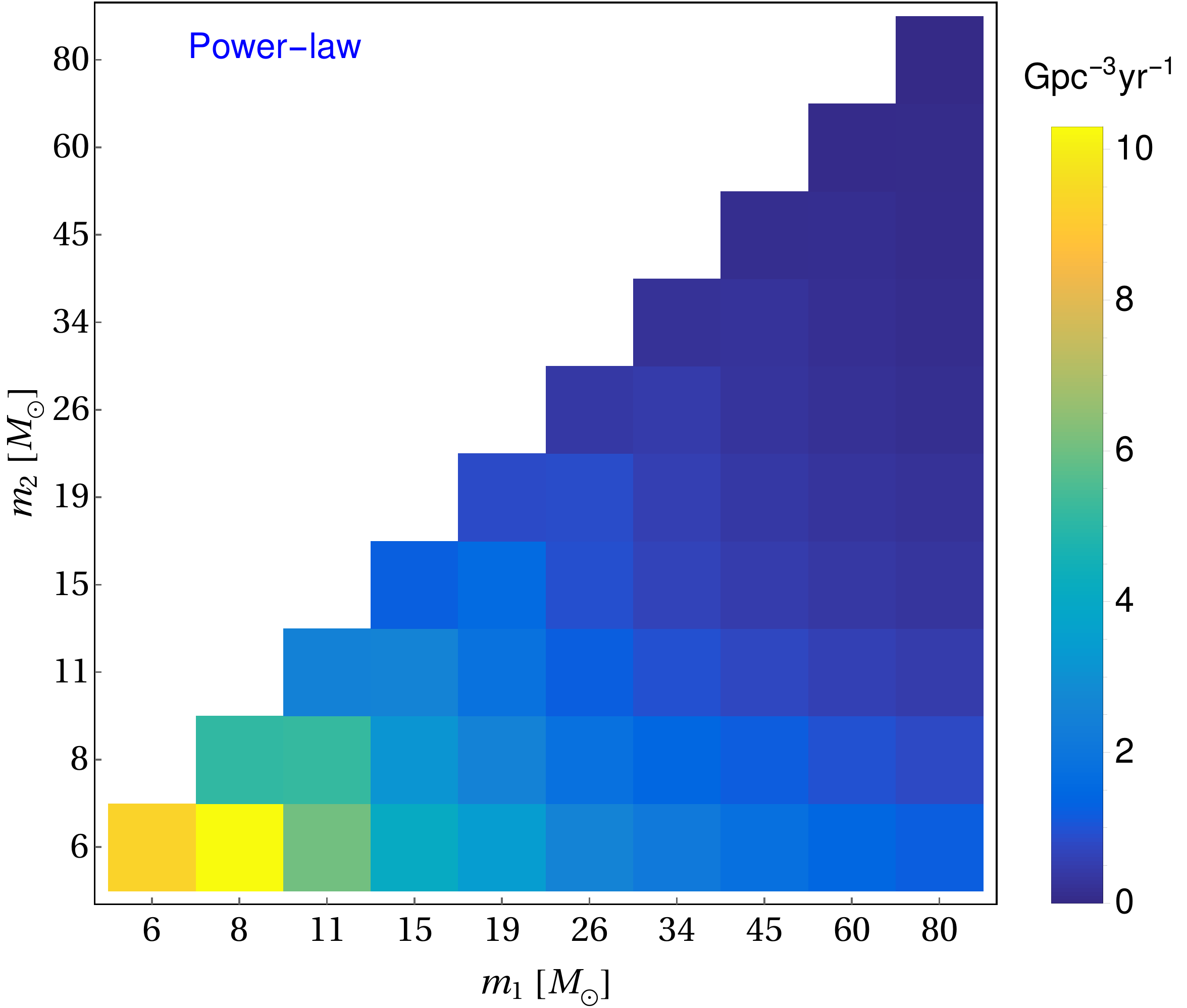}\\
\includegraphics[width = 0.48\textwidth]{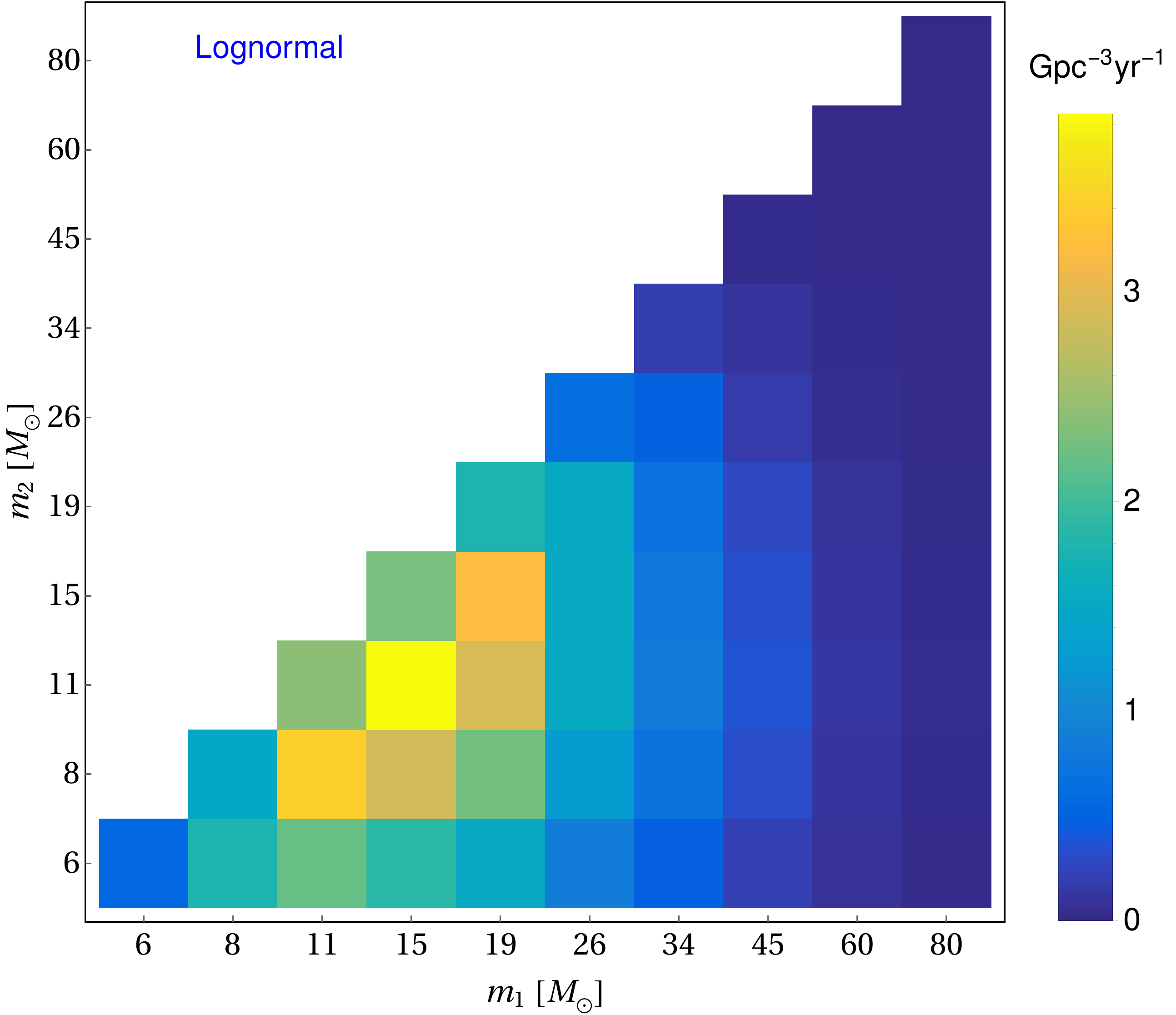}
\caption{\label{density}
The 2D merger rate distributions. The top and bottom panels correspond to the \textit{power-law} 
PDF ($M=5M_\odot$ and $\alpha=1.6$) with $\fpbh = 4.3\times 10^{-3}$
and \textit{lognormal} PDF ($m_c=15M_\odot$ and $\sigma=0.6$) with $\fpbh = 3.7\times 10^{-3}$, respectively.  
}
\end{figure}

Let's consider two typical PBH mass functions in literature. One takes the power-law form \cite{Carr:1975qj} as follows 
\m\label{power}
 P(m)\approx {\alpha-1\over M} \({m\over M}\)^{-\alpha}
\n
for $m\geq M$ and $\alpha>1$, and the other has a lognormal distribution \cite{Dolgov:1992pu}
\m\label{log}
 P(m) = \frac{1}{\sqrt{2 \pi} \sigma m} 
   \exp\(-\frac{\log^2(m/m_c)}{2 \sigma^2}\). 
\n
LIGO/VIRGO gives the merger rate for BBHs with $m_1,\ m_2\geq 5M_\odot$ and $m_1+m_2\leq 100M_\odot$ as $R_T=12\sim 213$ Gpc$^{-3}$ yr$^{-1}$ in \cite{Abbott:2017vtc}. Similar to \cite{Ali-Haimoud:2017rtz}, we take $\sigma_{\rm{eq}}\approx 0.005$. Fig.~\ref{merger} indicates that the merger rate constrained by LIGO/VIRGO can be explained by mergers of PBH binaries. 
Here, for simplicity, we take $M=5\Msun$ and $\alpha=1.6$ 
for the power-law PDF, and $m_c=15M_\odot$ and $\sigma=0.6$ for the lognormal PDF. Therefore LIGO/VIRGO implies that $1.4\times 10^{-3} \lesssim f_{\rm{pbh}}\lesssim 6.6\times 10^{-3}$ for the power-law PDF and $1.2\times 10^{-3} \lesssim f_{\rm{pbh}}\lesssim 5.7\times 10^{-3}$ for the lognormal PDF. Such an abundance of PBHs is consistent with current constraints from other observations \cite{Chen:2016pud,Green:2016xgy,Schutz:2016khr,Wang:2016ana,Gaggero:2016dpq,Ali-Haimoud:2016mbv,Blum:2016cjs,Horowitz:2016lib,Kuhnel:2017pwq,Inoue:2017csr,Carr:2017jsz,Green:2017qoa,Guo:2017njn,Zumalacarregui:2017qqd,Clesse:2016vqa}. 
In order to break the degeneracy for different PBH mass functions, we need more information. Keeping total merger rate $R_T=100$ Gpc$^{-3}$ yr$^{-1}$ fixed, we obtain $\fpbh = 4.3\times 10^{-3}$ for the power-law PDF and $\fpbh = 3.7\times 10^{-3}$ for the lognormal PDF respectively, and then plot the one-dimensional (1D) merger rate distribution (where we integrate over the mass of the lighter BH in the binary from $5M_\odot$ to the mass of heavier BH) in Fig.~\ref{mergerheavy}. 
We see that 1D merger rate distributions for the lognormal and power-law PDFs are quite different from each other even though both PDFs give the same total merger rate $R_T$. 
Furthermore, more information will be obtained in the two-dimensional (2D) merger rate distributions in  Fig.~\ref{density}. 


\begin{figure}[htbp!]
\centering
\includegraphics[width = 0.48\textwidth]{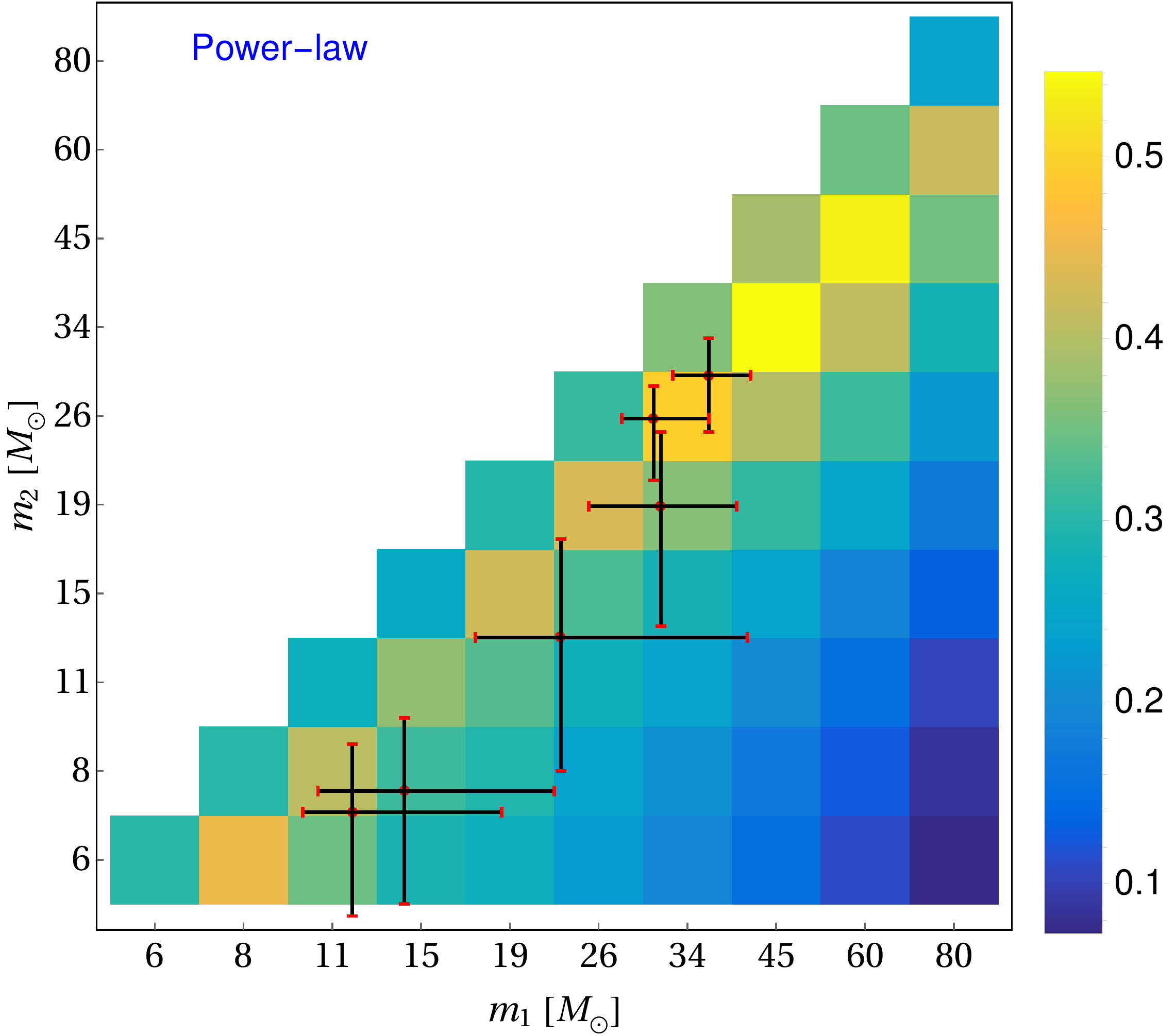}\\
\includegraphics[width = 0.48\textwidth]{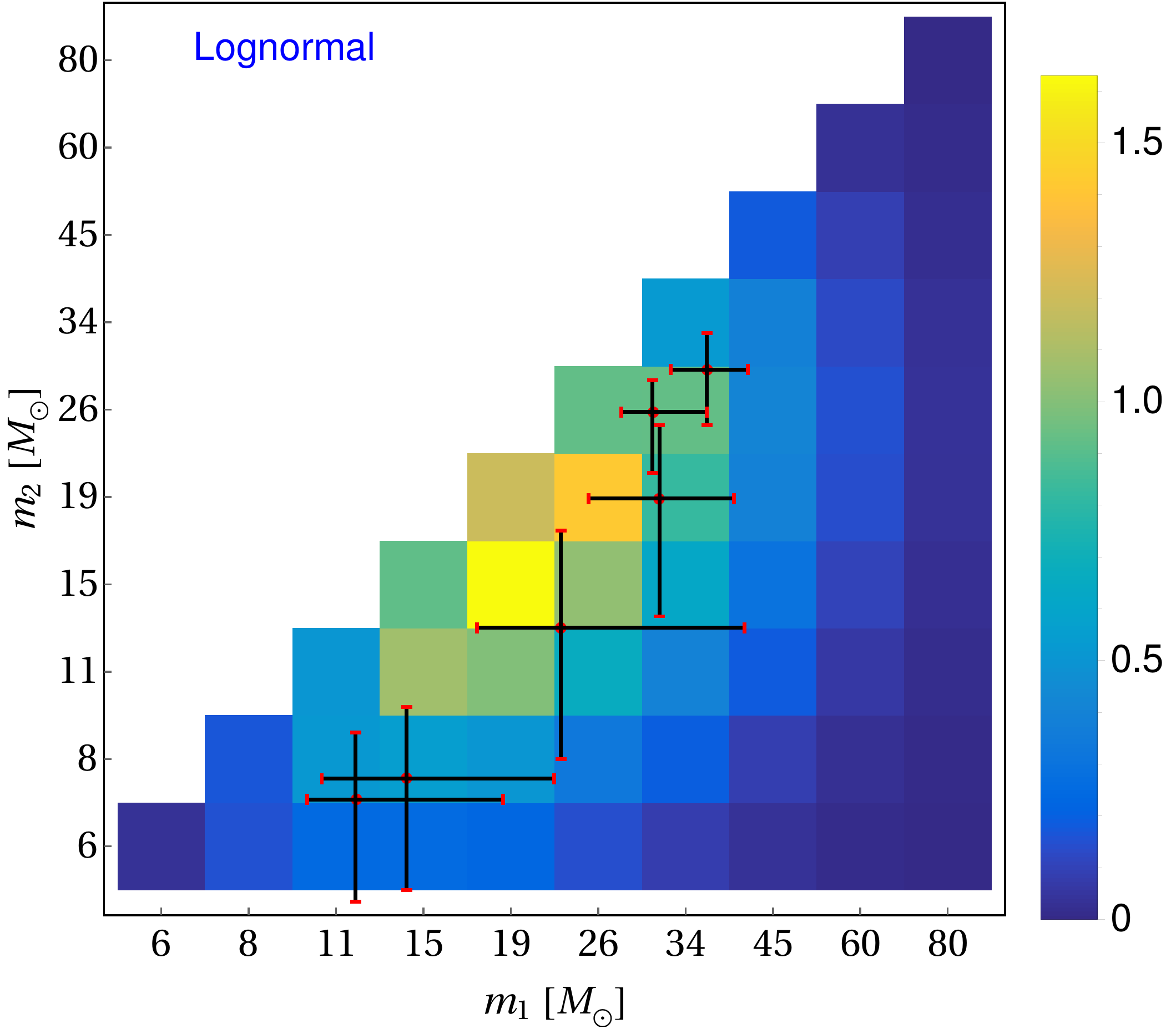}
\caption{\label{events}
The 2D distributions for $\Lambda$ [see Eq.~\eqref{lambda}],
along with the $6$ events detected by LIGO/VIRGO. 
The crosses indicate error bars for each event.
The top and bottom panels correspond to the \textit{power-law} 
PDF ($M=5M_\odot$ and $\alpha=1.6$) with $\fpbh = 4.3\times 10^{-3}$
and \textit{lognormal} PDF ($m_c=15M_\odot$ and $\sigma=0.6$) with $\fpbh = 3.7\times 10^{-3}$, respectively.  
}
\end{figure}

In order to compare to the events detected by LIGO, we need to take the sensitivity of LIGO into account. 
Since LIGO probes mergers approximately in the redshift range
$z \in [0,1]$, the expected number of triggers, $\Lambda$, is therefore 
estimated as \cite{Abbott:2016nhf,Abbott:2016drs,Abbott:2017iws,Kavanagh:2018ggo}
\e\label{lambda}
    \Lambda_{ij} = \int_{0}^{1} R_{ij}(z) \frac{d\VT}{dz} dz,
\q 
where $\VT$ is the averaged sensitive spacetime volume of LIGO,
which depends on the masses of the merging BBH.
We adopt the semi-analytical approximation presented in 
\cite{Abbott:2016nhf,Abbott:2016drs,Usman:2015kfa,Veitch:2014wba} to 
calculate $\VT$.
Here we assume LIGO's O1 and O2 runs share a common sensitive
volume, with observing time $48.6$ days for O1 \cite{TheLIGOScientific:2016pea}
and $117$ days for O2 \cite{TheLIGOScientific:2017qsa}.
Note that $\Lambda$ is not the mean number of confidently
detected binary-black-hole events, but the mean number of signals above
the chosen threshold \cite{Abbott:2016nhf}.
The 2D distributions of $\Lambda$, along with the $6$ events detected 
by LIGO/VIRGO, are then shown in Fig.~\ref{events}.
Since there are only few events available so far, we cannot make
a decisive conclusion for which PDF fits the data better.
As the data accumulating, we may finally pin down the shape of BH mass
function and nail down whether the lower and/or upper mass cutoffs
\cite{Kovetz:2016kpi,Fishbach:2017zga} for the
merging BH binaries do exist or not.

In this paper we work out the merger rate distribution of PBH binaries with a general mass function, by taking into account the torques by all other PBHs and the linear density perturbations. 
In \cite{Ali-Haimoud:2017rtz}, the effects of the tidal field from the smooth halo, the encountering with other PBHs, the baryon accretion and present-day halos were carefully investigated, and they concluded that all of these effects make no significant contributions to the overall merger rate. So it is reasonable for us to neglect these subdominant effects throughout our estimation as well. In addition, we find that the evolution of merger rate of PBH binaries goes like $t^{-34/37}$ which is quite different from that for the astrophysical black hole binaries. 

For the power-law and lognormal cases, the abundance of PBH is roughly
constrained to the range $10^{-3} \lesssim \fpbh \lesssim 10^{-2}$, 
which is consistent with other observations \cite{Chen:2016pud,Green:2016xgy,Schutz:2016khr,Wang:2016ana,Gaggero:2016dpq,Ali-Haimoud:2016mbv,Blum:2016cjs,Horowitz:2016lib,Kuhnel:2017pwq,Inoue:2017csr,Carr:2017jsz,Green:2017qoa,Guo:2017njn,Zumalacarregui:2017qqd,Clesse:2016vqa}. 
Our results hence confirm that the dominant fraction of DM should not 
originate from the stellar mass PBHs
\cite{Sasaki:2016jop,Ali-Haimoud:2017rtz,Raidal:2017mfl,Kocsis:2017yty}. 

A possible method to discriminate a PBH scenario from the others is to measure the merger rate distribution of BH binaries. In particular, we show that the 1D and 2D merger rate distributions are quite sensitive to the mass function of PBHs. In the near future, many more coalescences of BH binaries will be detected and provide the distribution of the binary parameters, and hence the mass function of PBH (or BH) could be reconstructed from the 1D and 2D merger rate distributions. It may finally help us to answer what is the origin of black holes detected by LIGO and VIRGO collaborations and how the black hole binaries are formed.

Acknowledgments. 
We thank the anonymous referee for valuable suggestions and comments.
We acknowledge the use of HPC Cluster of ITP-CAS. This work is supported by grants from NSFC (grant NO. 11335012, 11575271, 11690021, 11747601), Top-Notch Young Talents Program of China, and partly supported by the Strategic Priority Research Program of CAS and Key Research Program of Frontier Sciences of CAS.




\begin{thebibliography}{99}
\frenchspacing


\bibitem{Hawking:1971ei} 
  S.~Hawking,
  Mon.\ Not.\ Roy.\ Astron.\ Soc.\  {\bf 152}, 75 (1971).

\bibitem{Carr:1974nx} 
  B.~J.~Carr and S.~W.~Hawking,
  Mon.\ Not.\ Roy.\ Astron.\ Soc.\  {\bf 168}, 399 (1974).

\bibitem{Carr:1975qj} 
  B.~J.~Carr,
  Astrophys.\ J.\  {\bf 201}, 1 (1975).


\bibitem{Abbott:2016blz} 
  B.~P.~Abbott {\it et al.} [LIGO Scientific and Virgo Collaborations],
  Phys.\ Rev.\ Lett.\  {\bf 116}, no. 6, 061102 (2016)
  [arXiv:1602.03837 [gr-qc]].




\bibitem{Sasaki:2016jop} 
  M.~Sasaki, T.~Suyama, T.~Tanaka and S.~Yokoyama,
  Phys.\ Rev.\ Lett.\  {\bf 117}, no. 6, 061101 (2016)
  [arXiv:1603.08338 [astro-ph.CO]].

\bibitem{Bird:2016dcv} 
  S.~Bird, I.~Cholis, J.~B.~Mu–oz, Y.~Ali-Ha•moud, M.~Kamionkowski, E.~D.~Kovetz, A.~Raccanelli and A.~G.~Riess,
  Phys.\ Rev.\ Lett.\  {\bf 116}, no. 20, 201301 (2016)
  [arXiv:1603.00464 [astro-ph.CO]].

\bibitem{Abbott:2016nmj} 
  B.~P.~Abbott {\it et al.} [LIGO Scientific and Virgo Collaborations],
  Phys.\ Rev.\ Lett.\  {\bf 116}, no. 24, 241103 (2016)
  [arXiv:1606.04855 [gr-qc]].

\bibitem{Abbott:2017vtc} 
  B.~P.~Abbott {\it et al.} [LIGO Scientific and VIRGO Collaborations],
  Phys.\ Rev.\ Lett.\  {\bf 118}, no. 22, 221101 (2017)
  [arXiv:1706.01812 [gr-qc]].

\bibitem{Abbott:2017gyy} 
  B.~P.~Abbott {\it et al.} [LIGO Scientific and Virgo Collaborations],
  Astrophys.\ J.\  {\bf 851}, no. 2, L35 (2017)
  [arXiv:1711.05578 [astro-ph.HE]].

\bibitem{Abbott:2017oio} 
  B.~P.~Abbott {\it et al.} [LIGO Scientific and Virgo Collaborations],
  Phys.\ Rev.\ Lett.\  {\bf 119}, no. 14, 141101 (2017)
  [arXiv:1709.09660 [gr-qc]].

\bibitem{TheLIGOScientific:2016qqj} 
  B.~P.~Abbott {\it et al.} [LIGO Scientific and Virgo Collaborations],
  Phys.\ Rev.\ D {\bf 93}, no. 12, 122003 (2016)
  [arXiv:1602.03839 [gr-qc]].
  
\bibitem{TheLIGOScientific:2016pea} 
  B.~P.~Abbott {\it et al.} [LIGO Scientific and Virgo Collaborations],
  Phys.\ Rev.\ X {\bf 6}, no. 4, 041015 (2016)
  [arXiv:1606.04856 [gr-qc]].
  
\bibitem{Nakamura:1997sm} 
  T.~Nakamura, M.~Sasaki, T.~Tanaka and K.~S.~Thorne,
  Astrophys.\ J.\  {\bf 487}, L139 (1997)
  [astro-ph/9708060].    

\bibitem{Ali-Haimoud:2017rtz} 
  Y.~Ali-Ha•moud, E.~D.~Kovetz and M.~Kamionkowski,
  Phys.\ Rev.\ D {\bf 96}, no. 12, 123523 (2017)
  [arXiv:1709.06576 [astro-ph.CO]].


\bibitem{Nishikawa:2017chy} 
  H.~Nishikawa, E.~D.~Kovetz, M.~Kamionkowski and J.~Silk,
  arXiv:1708.08449 [astro-ph.CO].

\bibitem{Raidal:2017mfl} 
  M.~Raidal, V.~Vaskonen and H.~VeermŠe,
  JCAP {\bf 1709}, 037 (2017)
  [arXiv:1707.01480 [astro-ph.CO]].

\bibitem{Kocsis:2017yty} 
  B.~Kocsis, T.~Suyama, T.~Tanaka and S.~Yokoyama,
  arXiv:1709.09007 [astro-ph.CO].
  
\bibitem{Chandrasekhar:1943ws} 
  S.~Chandrasekhar,
  Rev.\ Mod.\ Phys.\  {\bf 15}, 1 (1943).  
  
\bibitem{Peters:1964zz} 
  P.~C.~Peters,
  Phys.\ Rev.\  {\bf 136}, B1224 (1964).

\bibitem{nt}
If we suppose that only the PBHs with mass not less than ${\rm{min}}(m_i,m_j)$ within the volume of $4\pi x^3/3$ disrupt the $m_i-m_j$ PBH pair, $n_T$ in Eq.~\eqref{dpX} should be modified to $n_T=f\rho_{\rm{eq}}\int_{{\rm{min}}(m_i,m_j)}^\infty P(m)/m dm$. 
    
\bibitem{xmu}
We check that the condition of ${X_*\over \mu}\cdot {4\pi \over 3}\langle x_{ij} \rangle^3 n_T\ll 1$ is satisfied in the examples in this paper. If ${X_*\over \mu}\cdot {4\pi \over 3}\langle x_{ij} \rangle^3 n_T\gtrsim 1$, one should use Eq.~\eqref{dP} to calculate the merger rate.


\bibitem{Dolgov:1992pu} 
  A.~Dolgov and J.~Silk,
  Phys.\ Rev.\ D {\bf 47}, 4244 (1993).



\bibitem{Chen:2016pud} 
  L.~Chen, Q.~G.~Huang and K.~Wang,
  JCAP {\bf 1612}, no. 12, 044 (2016)
  [arXiv:1608.02174 [astro-ph.CO]].

\bibitem{Green:2016xgy} 
  A.~M.~Green,
  Phys.\ Rev.\ D {\bf 94}, no. 6, 063530 (2016)
  [arXiv:1609.01143 [astro-ph.CO]].

\bibitem{Schutz:2016khr} 
  K.~Schutz and A.~Liu,
  Phys.\ Rev.\ D {\bf 95}, no. 2, 023002 (2017)
  [arXiv:1610.04234 [astro-ph.CO]].

\bibitem{Wang:2016ana} 
  S.~Wang, Y.~F.~Wang, Q.~G.~Huang and T.~G.~F.~Li,
  arXiv:1610.08725 [astro-ph.CO].

\bibitem{Gaggero:2016dpq} 
  D.~Gaggero, G.~Bertone, F.~Calore, R.~M.~T.~Connors, M.~Lovell, S.~Markoff and E.~Storm,
  Phys.\ Rev.\ Lett.\  {\bf 118}, no. 24, 241101 (2017)
  [arXiv:1612.00457 [astro-ph.HE]].

\bibitem{Ali-Haimoud:2016mbv} 
  Y.~Ali-Ha•moud and M.~Kamionkowski,
  Phys.\ Rev.\ D {\bf 95}, no. 4, 043534 (2017)
  [arXiv:1612.05644 [astro-ph.CO]].

\bibitem{Blum:2016cjs} 
  D.~Aloni, K.~Blum and R.~Flauger,
  JCAP {\bf 1705}, no. 05, 017 (2017)
  [arXiv:1612.06811 [astro-ph.CO]].
  
\bibitem{Horowitz:2016lib} 
  B.~Horowitz,
  arXiv:1612.07264 [astro-ph.CO].  
  
\bibitem{Kuhnel:2017pwq} 
  F.~KŸhnel and K.~Freese,
  Phys.\ Rev.\ D {\bf 95}, no. 8, 083508 (2017)
  [arXiv:1701.07223 [astro-ph.CO]].
  
\bibitem{Inoue:2017csr} 
  Y.~Inoue and A.~Kusenko,
  JCAP {\bf 1710}, no. 10, 034 (2017)
  [arXiv:1705.00791 [astro-ph.CO]].  
  
\bibitem{Carr:2017jsz} 
  B.~Carr, M.~Raidal, T.~Tenkanen, V.~Vaskonen and H.~VeermŠe,
  Phys.\ Rev.\ D {\bf 96}, no. 2, 023514 (2017)
  [arXiv:1705.05567 [astro-ph.CO]].  
    
\bibitem{Green:2017qoa} 
  A.~M.~Green,
  Phys.\ Rev.\ D {\bf 96}, no. 4, 043020 (2017)
  [arXiv:1705.10818 [astro-ph.CO]].
  
\bibitem{Guo:2017njn} 
  H.~K.~Guo, J.~Shu and Y.~Zhao,
  arXiv:1709.03500 [astro-ph.CO].  


\bibitem{Zumalacarregui:2017qqd} 
  M.~Zumalacarregui and U.~Seljak,
  arXiv:1712.02240 [astro-ph.CO].
  
\bibitem{Clesse:2016vqa} 
  S.~Clesse and J.~García-Bellido,
  Phys.\ Dark Univ.\  {\bf 15}, 142 (2017)
  [arXiv:1603.05234 [astro-ph.CO]].  

\bibitem{TheLIGOScientific:2017qsa} 
  B.~P.~Abbott {\it et al.} [LIGO Scientific and Virgo Collaborations],
  Phys.\ Rev.\ Lett.\  {\bf 119}, no. 16, 161101 (2017)
  [arXiv:1710.05832 [gr-qc]].

\bibitem{Abbott:2016nhf} 
  B.~P.~Abbott {\it et al.} [LIGO Scientific and Virgo Collaborations],
  Astrophys.\ J.\  {\bf 833}, no. 1, L1 (2016)
  [arXiv:1602.03842 [astro-ph.HE]]. 
     
\bibitem{Abbott:2016drs} 
  B.~P.~Abbott {\it et al.} [LIGO Scientific and Virgo Collaborations],
  Astrophys.\ J.\ Suppl.\  {\bf 227}, no. 2, 14 (2016)
  [arXiv:1606.03939 [astro-ph.HE]].

\bibitem{Abbott:2017iws} 
  B.~P.~Abbott {\it et al.} [LIGO Scientific and Virgo Collaborations],
  Phys.\ Rev.\ D {\bf 96}, no. 2, 022001 (2017)
  [arXiv:1704.04628 [gr-qc]].

\bibitem{Kavanagh:2018ggo} 
  B.~J.~Kavanagh, D.~Gaggero and G.~Bertone,
  arXiv:1805.09034 [astro-ph.CO].

\bibitem{Usman:2015kfa} 
  S.~A.~Usman {\it et al.},
  Class.\ Quant.\ Grav.\  {\bf 33}, no. 21, 215004 (2016)
  [arXiv:1508.02357 [gr-qc]].

\bibitem{Veitch:2014wba} 
  J.~Veitch {\it et al.},
  Phys.\ Rev.\ D {\bf 91}, no. 4, 042003 (2015)
  [arXiv:1409.7215 [gr-qc]].

\bibitem{Kovetz:2016kpi} 
  E.~D.~Kovetz, I.~Cholis, P.~C.~Breysse and M.~Kamionkowski,
  Phys.\ Rev.\ D {\bf 95}, no. 10, 103010 (2017)
  [arXiv:1611.01157 [astro-ph.CO]].

\bibitem{Fishbach:2017zga} 
  M.~Fishbach and D.~E.~Holz,
  Astrophys.\ J.\  {\bf 851}, no. 2, L25 (2017)
  [arXiv:1709.08584 [astro-ph.HE]].

\end{thebibliography}
\end{document}